\documentclass[arps,onecolumn,showpacs]{revtex4}

\usepackage{graphicx}
\usepackage{dcolumn}
\usepackage{bm}
\usepackage{hyperref}

\newcommand\lsim{\mathrel{\rlap{\lower4pt\hbox{\hskip1pt$\sim$}}
        \raise1pt\hbox{$<$}}}
\newcommand\gsim{\mathrel{\rlap{\lower4pt\hbox{\hskip1pt$\sim$}}
        \raise1pt\hbox{$>$}}}
\def\beq{\begin{equation}}
\def\eeq{\end{equation}}
\def\bea{\begin{eqnarray}}
\def\eea{\end{eqnarray}}

\def\gs{\mathrel{\raise1.16pt\hbox{$>$}\kern-7.0pt %
\lower3.06pt\hbox{{$\scriptstyle \sim$}}}}         %
\def\ls{\mathrel{\raise1.16pt\hbox{$<$}\kern-7.0pt %
\lower3.06pt\hbox{{$\scriptstyle \sim$}}}}         %

\begin{document}


\title{Unequal-Time Correlators for Cosmology}

\author{T. D. Kitching}
 \email{t.kitching@ucl.ac.uk}
 \affiliation{Mullard Space Science Laboratory, University College London, Holmbury St Mary, Surrey RH5 6NT, UK}
\author{A. F. Heavens}%
 \email{a.heavens@imperial.ac.uk}
\affiliation{ICIC, Astrophysics, Imperial College, Blackett Laboratory, Prince Consort Road, London SW7 2AZ, UK}%

\date{\today}

\begin{abstract}
Measurements of the power spectrum from large-scale structure surveys have to date assumed 
an equal-time approximation, where the full cross-correlation power spectrum of the matter density field evaluated at different times (or distances) has 
been approximated either by the power spectrum at a fixed time, or in an improved fashion, by a geometric mean $P(k; r_1, r_2)=[P(k; r_1) P(k; r_2)]^{1/2}$.
In this paper we investigate the
expected impact of the geometric mean ansatz, and present an application in assessing the impact on weak gravitational lensing
cosmological parameter inference, using a perturbative unequal-time correlator.  As one might expect, we find that the impact of this assumption is greatest at large separations
in redshift $\Delta z\gs 0.3$ where the change in the amplitude of the matter power spectrum can be as much as
$10$ percent for $k\gs 5h$Mpc$^{-1}$.  However, of more concern is that the corrections for small separations, where the clustering is not close to zero, may not be negligibly small. In particular, we find that
for a Euclid- or LSST-like weak lensing experiment the assumption of equal-time correlators may result in biased predictions of
the cosmic shear power spectrum, and that the impact is strongly dependent on the amplitude of the intrinsic alignment signal. 
To compute unequal-time correlations to sufficient accuracy will require advances 
in either perturbation theory to high $k$-modes, or extensive use of simulations.
\end{abstract}

\pacs{Valid PACS appear here}
\maketitle


\section{Introduction}
\label{Introduction}
We consider evolving scalar fields in cosmology. We define an evolving three-dimensional random field 
$A({\bf x}; r(t))$, where ${\bf x}$ labels a local three dimensional coordinate defined at a time $t$; 
$t$ equivalently labels a coordinate distance $r(t)$ and we may use these interchangeably. 
For such an evolving three-dimensional random field we can choose a time coordinate $t$ and define a 
local plane wave Fourier transform as 
\beq
A({\bf k}; r(t)) = \int  d^3{\bf x}~  e^{-i{\bf k\cdot x}} A({\bf x}; r(t)),  
 \eeq
where $k$ is a three dimensional wavenumber, and we use the notation of \cite{2008PhRvD..78l3506L}.  Even this equation involves a subtlety; we observe on the past light cone, so measurements are made not at a single time slice.  In writing this, we assume that we evolve the fields to a common time slice, and we assume in this paper that the integration volume is small enough that the evolution is small enough that it can be done essentially perfectly.   This then ensures that universal homogeneity is preserved in these local fields labelled with $t$.

This field considered could be, for instance, the density fluctuation $\delta\rho({\bf x})/\rho$, or the 
Newtonian potential $\Phi({\bf x})$ defined at time $t_i$. 
For any two fields each with a time label, the cross-correlation power spectrum, 
which is assumed to be isotropic, is defined by
\beq
\label{Pkexact}
\langle A({\bf k}; r(t_1))A^*({\bf k'}; r(t_2))\rangle = (2\pi)^3 \delta^3({\bf k-k'}) P(k; r(t_1), r(t_2)),
\eeq
and we have exploited the homogeneity discussed above. From here we will use the notation $r_i=r(t_i)$ for clarity. 
The projection of a field $A$ on the sky is defined using projection kernels $F_A$ such that 
\beq
\tilde{A}({\bf \hat{n}}; r) \,=\, \int_0^{r} dr_1~F_A(r,r_1) A(r_1{\bf \hat{n}}; r_1),
\eeq
where $r$ is a surface that labels the projection and characterises the time when the 
projection is performed, and ${\bf \hat{n}}$ represents direction on the sky.  A simple example of $F_A(r,r_1)$ is a 
Gaussian centred on $r$ with a certain width. Expanding $\tilde{A}$
in terms of spherical harmonics, the angular cross-power spectrum of two different projections, $C^{AB}_\ell(r, r')$ is defined as
\bea
\label{Clexact}
C_\ell^{AB}(r, r')&\equiv& \langle\tilde{A}_{\ell m}(r)\tilde{B}^*_{\ell m}(r')\rangle\nonumber\\
&=& \int_0^{r} dr_1 \int_0^{r} dr_2 F_A(r,r_1) F_B(r',r_2) 
\int \frac{ d^3{\bf k}}{(2\pi)^3} P(k; r_1, r_2) (4\pi)^2 j_{\ell}(kr_1)j_{\ell}(kr_2)Y_{\ell m}({\bf \hat{k}})Y^*_{\ell m}({\bf \hat{k}})\nonumber\\
&=& \int_0^{r} dr_1 \int_0^{r'} dr_2 F_A(r,r_1) F_B(r',r_2)
\int \frac{ 2 dk k^2}{\pi} P(k; r_1, r_2) j_{\ell}(kr_1)j_{\ell}(kr_2)
\eea
where $j_{\ell}$ are spherical Bessel functions of order $\ell$ and $Y_{\ell m}$ are spherical harmonics.
Here we consider the case that two fields $A$ and $B$ are different projections of the same underlying field and hence probe the same underlying power spectrum 
$P(k; r_1, r_2)$ but via different projection kernels $F_A(r,r_1)$ and $F_B(r,r_2)$ respectively. This is easily generalised to different fields, in which case the cross-power spectrum is involved.

In the full expression of equation (\ref{Clexact}) 
the power spectrum $P(k; r_1, r_2)$ is that between two different, unequal, times. We refer to 
$P(k; r_1, r_2)$ as the unequal-time correlator. In cosmology however underlying 
power spectra are usually only expressed as a function of $k$ and a single time, i.e., $P(k;r)$,  encoding the Fourier 
correlations at a given comoving distance, but not correctly describing the correlations between comoving distances. 

To overcome this issue a mixed equal-time approximation was introduced in cosmology \cite{2005PhRvD..72b3516C}, where the geometric mean of the two equal-time power spectra has been assumed: 
$P(k; r_1, r_2)\simeq [P(k; r_1)P(k; r_2)]^{1/2}$.
This approximation is justified by assuming that the correlation 
of the underlying (matter overdensity) field is restricted to small-scales in cosmology, and that over such scales the 
look-back time is approximately equal ($r_1\simeq r_2$) therefore either $P(k; r_1)$ or $P(k; r_2)$ could be used instead 
of $P(k; r_1, r_2)$. The geometric mean approximation is then used as an algebraic convenience such that the integrals in 
equation (\ref{Clexact}) can be separated. A further justification is given that the Bessel function integrals 
may reduce to a delta-function for $r_1\not=r_2$ for large $kr$. These assumptions are tested in this paper. 

\section{Unequal-time power spectra}
\label{Unequal-time power spectra}
To compute the power spectrum $P(k; r_1, r_2)$ we need to refer to the correlation between the underlying 
fields in equation (\ref{Pkexact}). In the cosmological context the underlying field of interest is often the matter 
overdensity $\delta({\bf x},t)=[\rho({\bf x},t)-\bar\rho]/\bar\rho$; where $\rho({\bf x},t)$ is the matter 
density at a position ${\bf x}$ and comoving time $t$, and $\bar\rho$ is the mean matter density. This can be 
expanded \cite{1998MNRAS.301..797H,2006ApJ...651..619J} 
perturbatively in terms of a growth factor $D(t)$, that is independent of scale, 
and a transfer function 
\beq 
\label{delta}
\delta({\bf k},t)=\sum_{n=1}^{\infty} D^n(t)f_n({\bf k}),
\eeq
where $\delta({\bf k},t)$ is the three-dimensional Fourier transform of $\delta({\bf x},t)$. 
$f_n({\bf k})$ involves products and integrals of $n$ terms of $\delta_{\rm L}({\bf k},t)$ 
-- which is the density field that can be computed from linear gravitational 
theory \citep{1998MNRAS.301..797H}. 
For $n>3$ such terms become  complicated and numerically challenging to compute. 

We can now take correlations of equation (\ref{delta}) in both the unequal-time and the equal-time cases.
For the unequal-time correlation we have
\bea
\label{puetc}
P^{\rm UETC}(k; r, r')&=&\langle \delta({\bf k},t)\delta^*({\bf k},t')\rangle\nonumber\\
&=&\sum_{n=1}^{\infty}\sum_{m=1}^{\infty} D^n(t)D^m(t')\langle f_n({\bf k})f^*_m({\bf k})\rangle\nonumber\\
&=&\sum_{n=1}^{\infty}\sum_{m=1}^{\infty} D^n(t)D^m(t') P_{nm}(k)\nonumber\\
&=&D(t)D(t')P_{11}(k)+D^2(t)D^2(t')P_{22}(k)+[D^3(t)D(t')+D(t)D^3(t')]P_{13}(k)+\dots
\eea
where $P_{nm}(k)$ are the power spectra corresponding to the perturbatively expanded $\delta({\bf k},t)$ 
at order $nm$, see e.g. \cite{2006ApJ...651..619J}. $D^n(t)$ is the 
linear growth factor at comoving time $t$ to the $n^{\rm th}$ power. 
In the fourth line we expand this expression to include all terms up to $P_{13}(k)$.

The equal-time case can be written as 
\bea
\label{petc}
P^{\rm ETC}(k; r(t))&=&\langle \delta({\bf k},t)\delta^*({\bf k},t)\rangle\nonumber\\
&=&\sum_{n=1}^{\infty}\sum_{m=1}^{\infty} D^n(t)D^m(t)\langle f_n({\bf k})f^*_m({\bf k})\rangle\nonumber\\
&=&\sum_{n=1}^{\infty}\sum_{m=1}^{\infty} D^n(t)D^m(t) P_{nm}(k)\nonumber\\
&=&D^2(t)P_{11}(k)+D^4(t)P_{22}(k)+2D^4(t)P_{13}(k)+\dots
\eea
where the functions $D$ and $P_{nm}$ are the same as in equation (\ref{puetc}). 

\begin{figure}
\begin{tabular}{cc}
\includegraphics[width=0.5\textwidth]{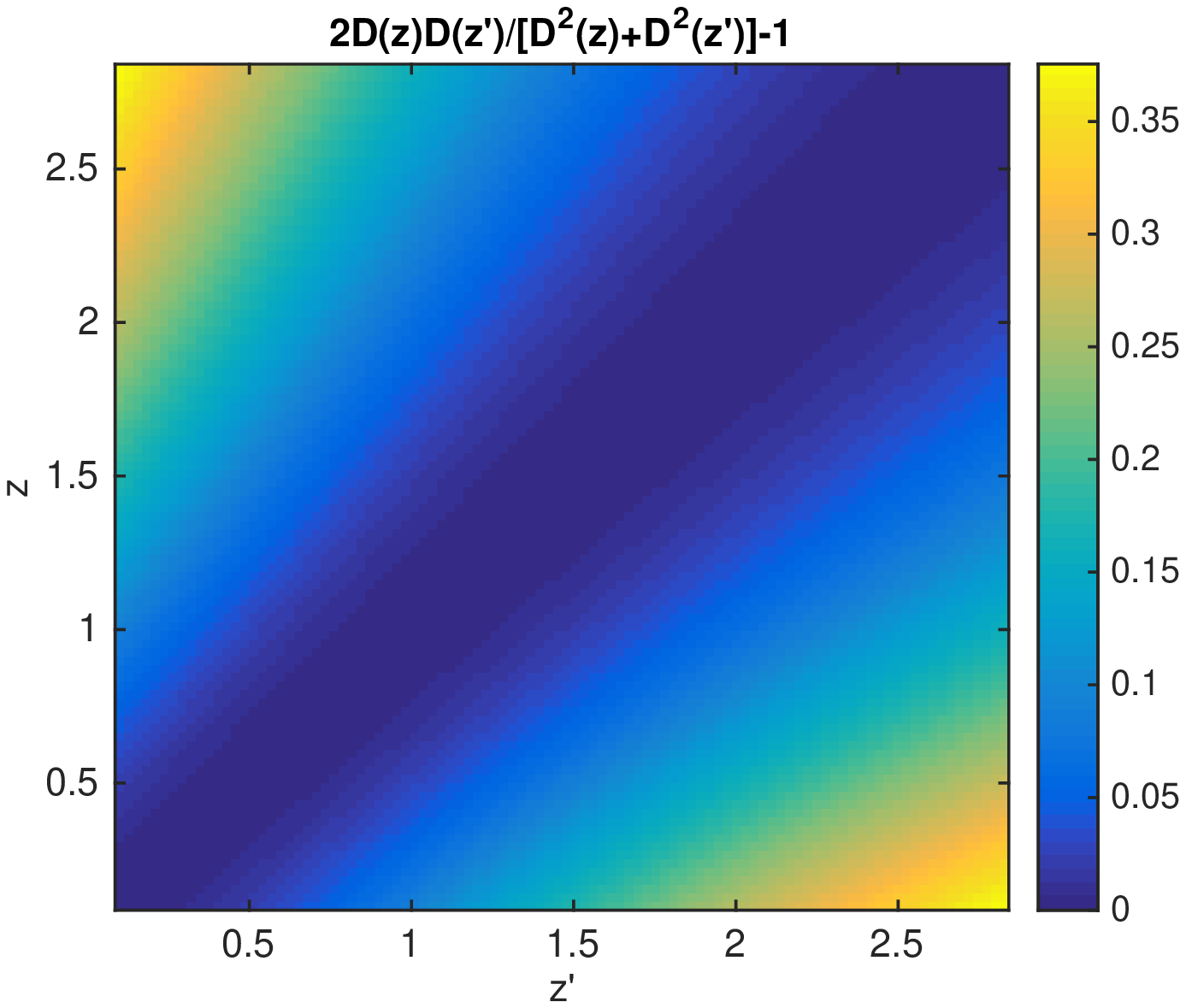}& \includegraphics[width=0.5\textwidth]{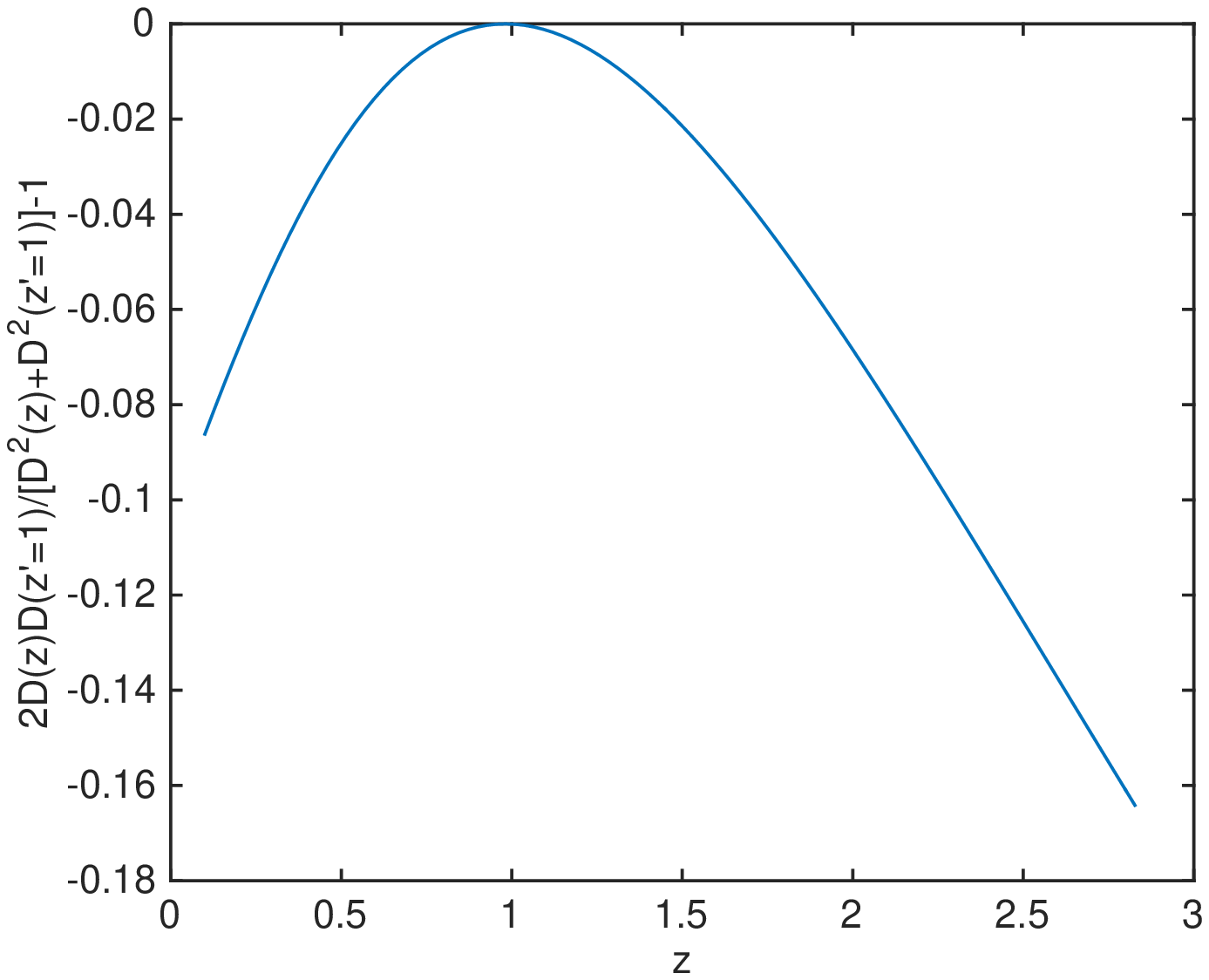}
\end{tabular}
\caption{Left Panel: The fractional difference between $2D(z)D(z')$ and 
$D^2(z)+D^2(z')$, where $D(z)$ is the linear growth factor, for a $\Lambda$CDM 
cosmology with parameters set at the Planck \cite{2016A&A...594A..13P} maximum likelihood values. Right Panel: 
Is a cross-section through the left panel plot for $z'=1$.}
\label{dplot}
\end{figure}

The geometric mean equal-time ansatz is that $[P(k; r, r')]^2=P^{\rm ETC}(k; r)P^{\rm ETC}(k; r')$. By squaring 
equation (\ref{puetc}) and multiplying equation (\ref{petc}) as required the differences between the two cases become clear 
\bea 
[P^{\rm UETC}(k; r, r')]^2&=&
D^2(t)D^2(t')P^2_{11}(k)+
D^4(t)D^4(t')P^2_{22}(k)+
[D^2(t)+D^2(t')]D^2(t)D^2(t)P^2_{13}(k)\nonumber\\
&+&
2D^3(t)D^3(t')P_{11}(k)P_{22}(k)+
2D^2(t)D^2(t')[D^2(t)+D^2(t')]P_{11}(k)P_{13}(k)\nonumber\\
&+&
2D^3(t)D^3(t')[D^2(t)+D^2(t')]P_{22}(k)P_{13}(k)
\eea
and 
\bea
P^{\rm ETC}(k; r)P^{\rm ETC}(k; r')&=&
D^2(t)D^2(t')P^2_{11}(k)+
D^4(t)D^4(t')P^2_{22}(k)+
4D^4(t)D^4(t)P^2_{13}(k)\nonumber\\
&+&
D^2(t)D^2(t')[D^2(t)+D^2(t')]P_{11}(k)P_{22}(k)+
2D^2(t)D^2(t')[D^2(t)+D^2(t')]P_{11}(k)P_{13}(k)\nonumber\\
&+&
4D^4(t)D^4(t')P_{22}(k)P_{13}(k), 
\eea
where each term can be compared.  We see that many terms are in common, but a few coincide only if
$D^2(t)+D^2(t')=2D(t)D(t')$, which evidently only holds when $D(t)=D(t')$.  
From here we will label comoving time with redshift $z$, where the assumption of a 
cosmology is implicit. In Figure \ref{dplot} we show the magnitude of this 
approximation as a function of redshift for a concordance cosmology.
\begin{figure}
\begin{tabular}{cc}
\includegraphics[width=0.5\textwidth]{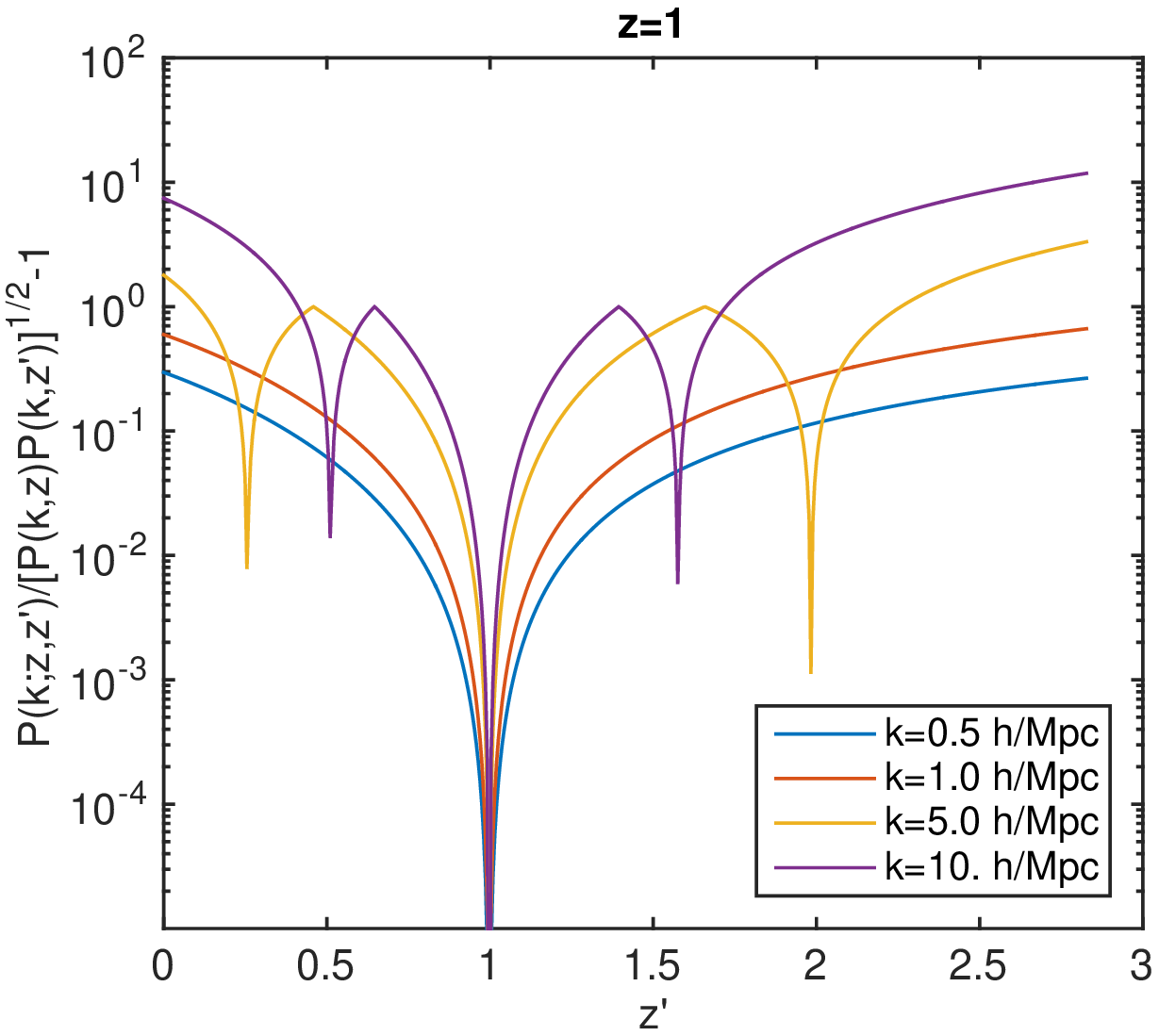}& \includegraphics[width=0.5\textwidth]{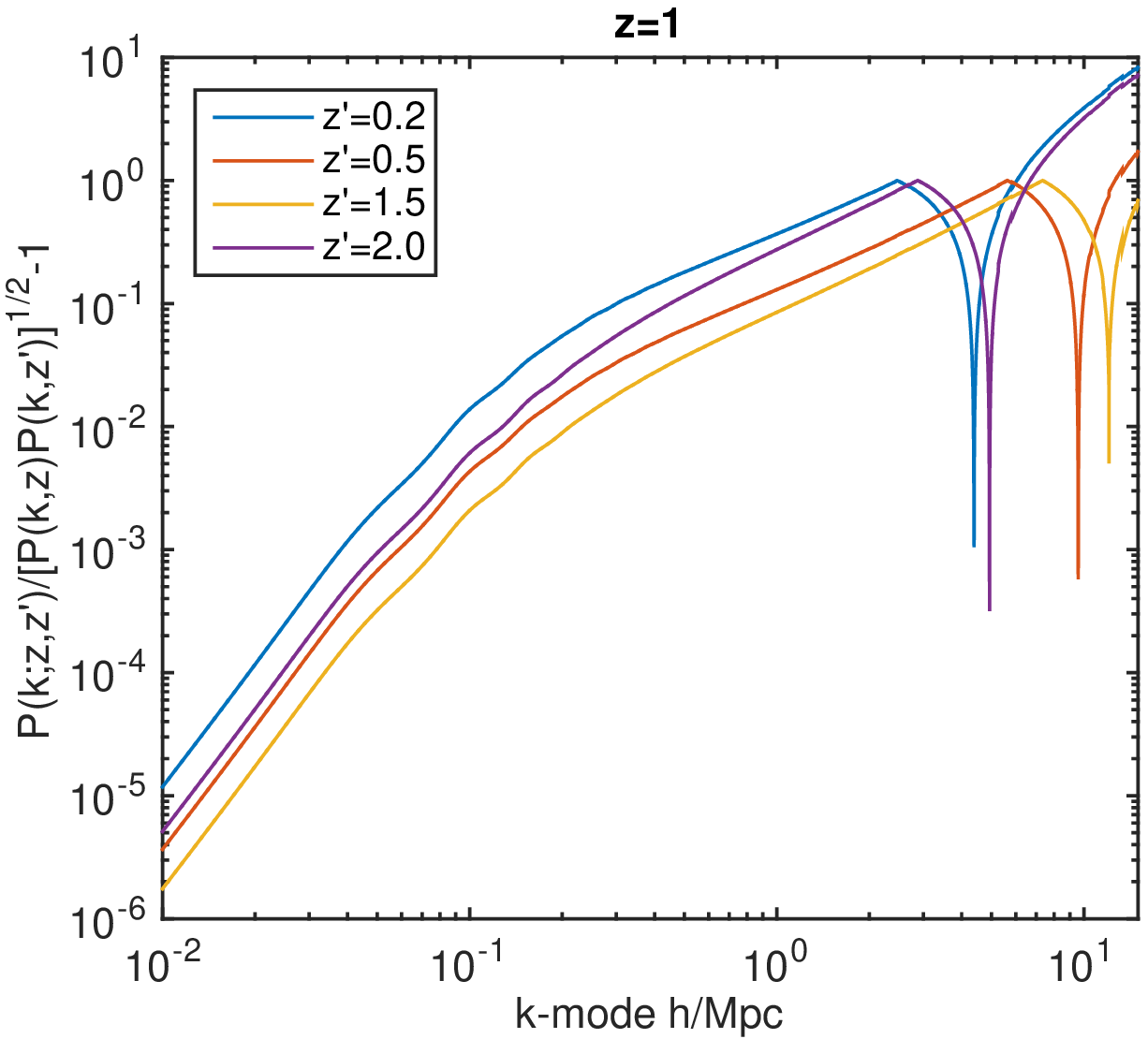}
\end{tabular}
\caption{The fractional difference between 
$P^{\rm UETC}(k; r, r')$ and $[P^{\rm ETC}(k; r)P^{\rm ETC}(k; r')]^{1/2}$ and
for a $\Lambda$CDM
cosmology with parameters set at the Planck \cite{2016A&A...594A..13P} maximum likelihood values. In the left panel 
we set $z=1$ and show the variation as a function of $z'$ for several $k$-modes . In the right panel
we set $z=1$ and show the variation as a function of $k$-mode for several $z'$ values.}
\label{pkplot}
\end{figure}
In Figure \ref{pkplot} we show $P^{\rm UETC}(k; r, r')$ and $[P^{\rm ETC}(k; r)P^{\rm ETC}(k; r')]^{1/2}$ 
calculated using equations (\ref{puetc}) and 
(\ref{petc})\footnote{For these calculations we use {\tt Cosmolopy} \url{http://roban.github.io/CosmoloPy/docAPI/cosmolopy.perturbation-module.html} to compute the growth factor, and the $3^{\rm rd}$ order perturbation code 
for $P_{11}$, $P_{22}$, and $P_{13}$ available from E. Komatsu 
here \url{http://wwwmpa.mpa-garching.mpg.de/\~komatsu/CRL/powerspectrum/density3pt/pkd/}.}. We
see that the impact of the equal-time approximation is largest at large redshift separations and increases 
in amplitude at small-scales where the $P_{22}(k)$ and $P_{13}(k)$ become dominant.
For separations in redshift $|z-z'|\gs 0.3$ the change in the amplitude of the matter power spectrum can be as much as
$10\%$ for $k\gs 5h$Mpc$^{-1}$. However, it is not clear yet whether the large deviations at large $|z-z'|$ are important, since the correlation of the fields is likely to be very small here. We will return to this later.

An alternative formulation of an unequal-time correlation is to use 
an `eikonal' phase 
\cite{1996ApJ...456...43J,2012PhRvD..85f3509B,2013PhRvD..87d3530B,2015JCAP...02..026B,2015JCAP...11..032G} where the 
matter overdensity perturbations can be written like \cite{2015JCAP...02..026B}
\beq 
\delta({\bf k},t)\simeq {\rm exp}\left[\int^t {\rm d}t'\int \frac{{\rm d}^3{\bf k'}}{(2\pi)^3}\frac{{\bf k}.{\bf k'}}{k'^2}
\delta_{\rm L}({\bf k},t')\right]\times \delta_{\rm S}({\bf k},t),
\eeq
where $\delta_{\rm S}({\bf k},r)$ represent fluctuations on short scales, and $\delta_{\rm L}({\bf k},r)$ are linear-scale perturbations; 
this equation assumes a growing mode only. 
The unequal-time power spectrum is then expressed as the equal-time version 
multiplied by exponential damping term that depends on the 
separation in time; see for example \cite{2015JCAP...02..026B} section 2.2. 
In this case unequal-time correlations are due to the mixing of long wavelength (or `soft') modes with 
shorter wavelength modes. It is found that for equal times these have no impact on the amplitude of the 
power spectrum, which is consistent 
with the conclusion we find in this Section, but that for unequal-times there can be 
significant changes \cite{2012PhRvD..85f3509B}. 
In \cite{2014JCAP...02..051C} it is shown how to link the eikonal phase representation to a perturbation approach 
similar to that used here. \cite{2015JCAP...02..026B,2015JCAP...11..032G} derive power spectrum 
and bispectrum consistency conditions for the equal and unequal-time cases using this formalism.  

\subsection{Bessel/Lommel Effects}
Despite the amplitude of the underlying power spectrum being different in the unequal-time and equal-time 
cases the integration over two Bessel functions in equation (\ref{Clexact}) may be expected to 
down-weight the impact of such an effect for a projected field. 
Indeed when integrating over multiple Bessel functions an orthogonality relation holds 
\bea 
\int_0^R {\rm d}r r^2 j_{\ell}(k_n r)j_{\ell}(k_m r)=j_{\ell+1}(k_n R)\delta^K_{nm}
\eea
where $k_nr$ are the arguments corresponding to zeros of the spherical Bessel functions\cite{watson1995treatise,abramowitz1964handbook}. 
However for the general case, where non-zero parts of the spherical 
Bessel function are integrated over, this orthogonality relation does not hold. There have been several investigations 
into the regimes where this expression is applicable for cosmology \cite{1995MNRAS.272..885F, 1995MNRAS.275..483H}.
In the case that there are such integrals over $r$ such a relation can be applied in the derivation of quantities for 
cosmology. However in the case of projected fields, that we investigate here, this is not appropriate 
because the equivalent expression that 
appears in equation (\ref{Clexact}) involves an integral over $k$-mode. In the constant power spectrum case, 
and an upper $k$-mode limit of $K$, we would have expressions like 
\bea
\int_0^{K} {\rm d}k k^2  j_{\ell}(kr_n)j_{\ell}(kr_m)=j_{\ell+1}(K r_n)\delta^K_{nm},
\eea
which is only applicable for a discretely sampled field in comoving distance $r_n$ such 
that $kr_n$ are zeros of the spherical Bessel functions, this is clearly not the case. 
In \cite{2005PhRvD..72b3516C,Baddour:10,2012A&A...540A..60L} the case of $K\rightarrow \infty$ is 
discussed which results in 
\bea
\label{edk}
\int_0^{\infty} {\rm d}k k^2  j_{\ell}(kr)j_{\ell}(kr')=\frac{\pi}{2k^2}\delta^D(r-r');
\eea
but this is not a practical case for cosmology, and furthermore the full expression for $K < \infty$ 
does not converge to limit as we will discuss. 

\begin{figure}
\begin{tabular}{cc}
\includegraphics[width=0.5\textwidth]{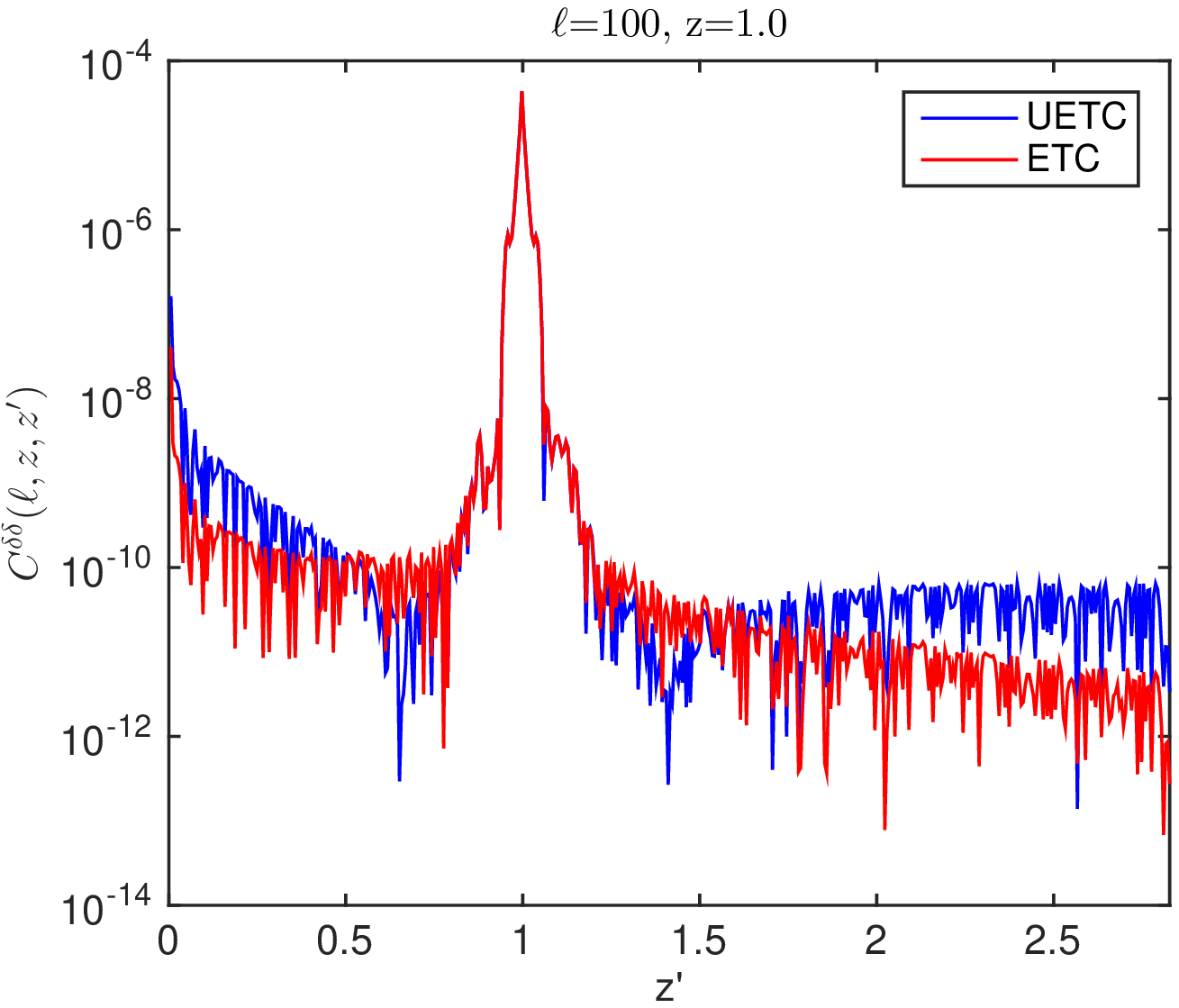}& \includegraphics[width=0.5\textwidth]{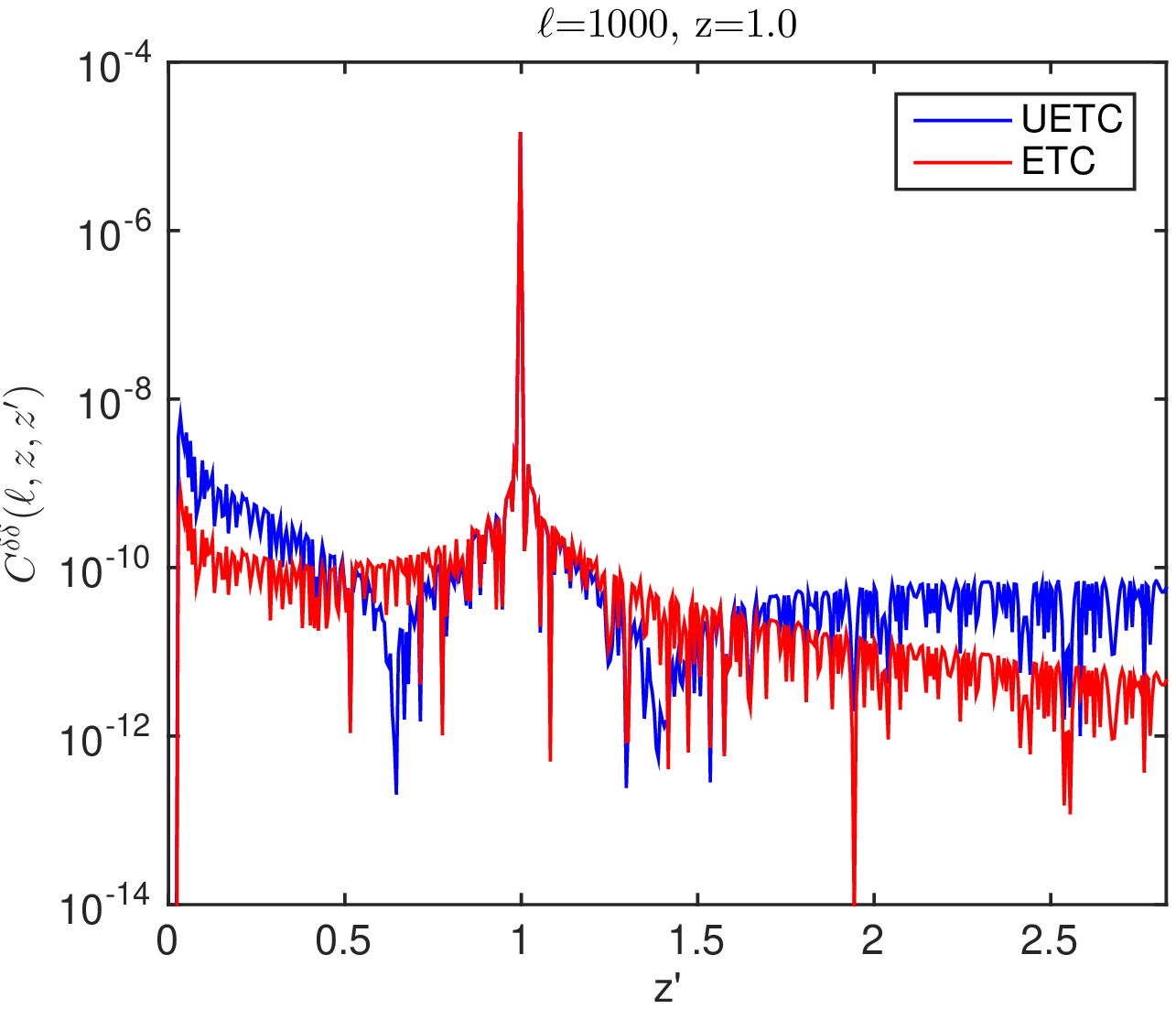}\\
\end{tabular}
\caption{The functions $C^{\delta\delta, {\rm UETC}}_{\ell}(r_1,r_2)$ (blue) and 
$C^{\delta\delta, {\rm ETC}}_{\ell}(r_1,r_2)$ (red) for several different $\ell$-modes and 
redshift values defined equations (\ref{qeq}). The title of each plot defines the $\ell$-mode and $z$ value 
used.}
\label{qplot}
\end{figure}

\begin{figure}
\begin{tabular}{cc}
\includegraphics[width=0.5\textwidth]{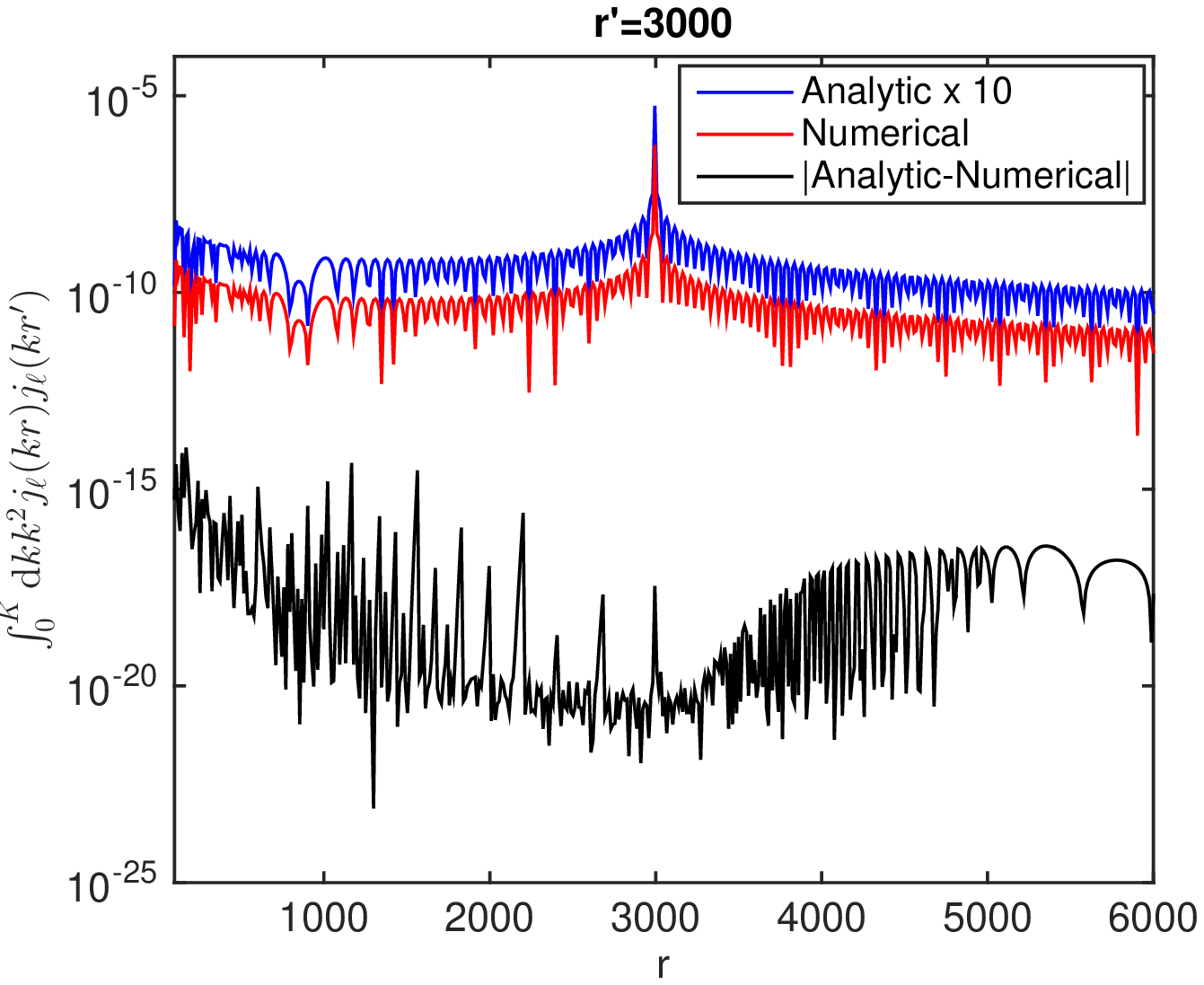}& \includegraphics[width=0.5\textwidth]{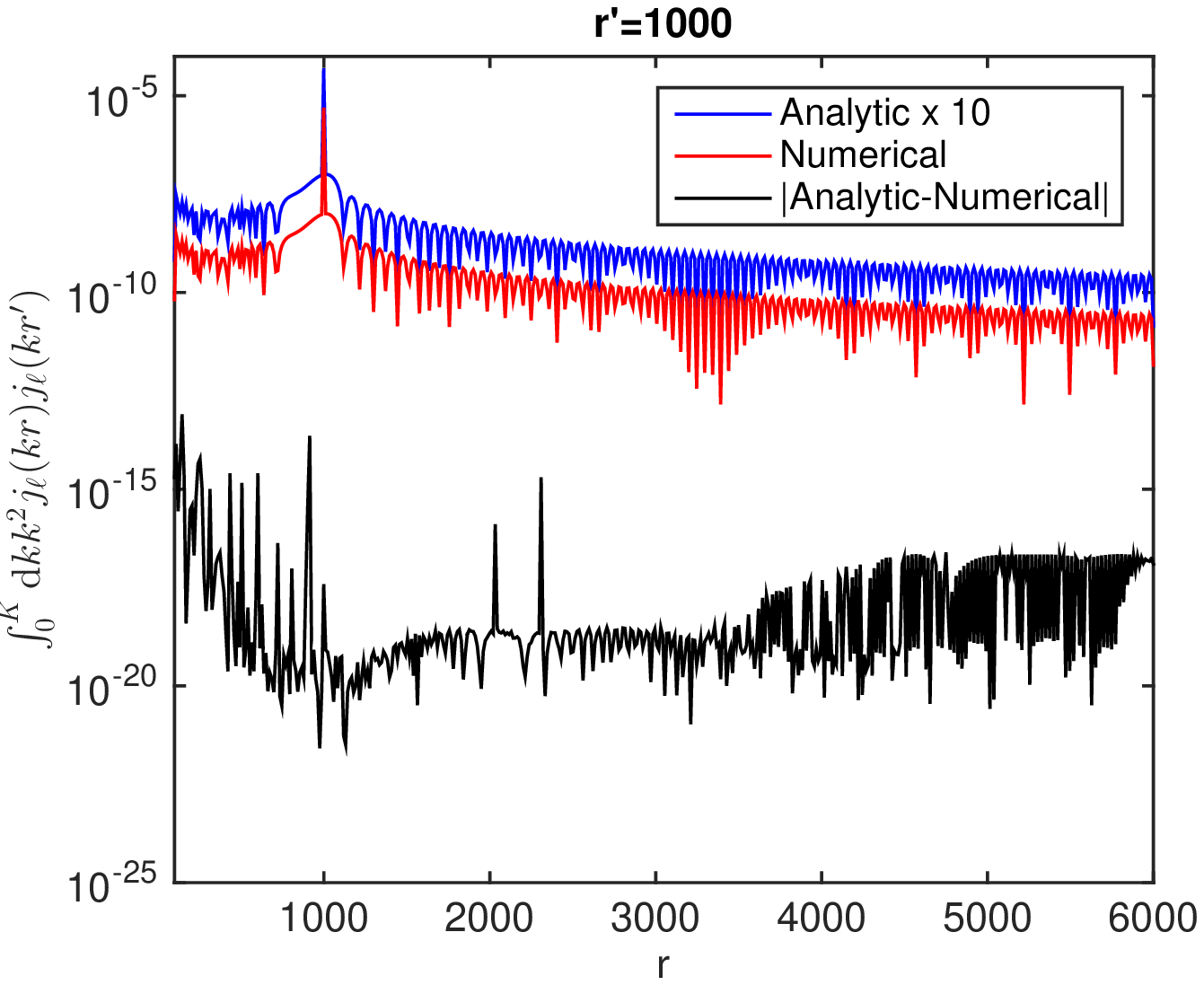}
\end{tabular}
\caption{Left Panel: Equation (\ref{lommel}) as a function of $r'$ for $r=3000$. The red line is calculated 
using numerical integration, and the blue line is the analytic solution mutlplied by ten so that the functional form can be distinguished. The black line shows the absolute difference between the two. Right Panel: Is the same as the left except $r=1000$. Both plots are for $\ell=1000$ and $K=10h$Mpc$^{-1}$.}
\label{lplot}
\end{figure}

To investigate the effect of the Bessel integrations we explore the functions 
\bea 
\label{qeq}
C^{\delta\delta, {\rm UETC}}_{\ell}(r_1,r_2)&=&\int \frac{ 2 dk k^2}{\pi} P^{\rm UETC}(k; r_1, r_2) j_{\ell}(kr_1)j_{\ell}(kr_2)\nonumber\\
C^{\delta\delta, {\rm ETC}}_{\ell}(r_1,r_2)&=&\int \frac{ 2 dk k^2}{\pi} [P^{\rm ETC}(k; r_1)P^{\rm ETC}(k,r_2)]^{1/2} j_{\ell}(kr_1)j_{\ell}(kr_2),
\eea
for the unequal-time and equal-time cases respectively. 
These enter in equation (\ref{Clexact}), and similarly for the equal-time case, and are 
then integrated over with the projection kernels $F_A(r,r_1)$ to generate the projected power spectra, 
\beq
\label{qeq2}
C^{\rm AB}_\ell(r, r')=\int_0^{r} dr_1 \int_0^{r'} dr_2 F_A(r,r_1) F_B(r',r_2)C^{\delta\delta}_{\ell}(r_1,r_2).
\eeq
Equations (\ref{qeq}) are also equal to the projected power spectra $C_\ell(r_A, r_B)$ when the 
projection kernels are delta-functions e.g. $\delta^D(r-r_1)$. 
We assume the same $P(k; r_1, r_2)$ as in Section \ref{Unequal-time power spectra}, where the integration 
is over range $10^{-3}\leq k \leq 15 h$Mpc$^{-1}$\footnote{We use the {\tt Matlab} integration 
routine {\tt quadgk} with a relative tolerance of $10^{-20}$ and an absolute tolerance of $10^{-14}$, 
that are sufficient for numerical convergence, see e.g. Figure \ref{lplot}.}. In Figure \ref{qplot} we show the 
functions for several different $\ell$-modes and redshift values. 
It can be seen that the functional form of $C^{\delta\delta}_{\ell}(r_1,r_2)$ is sharply peaked about 
$r_1=r_2$ but has a non-zero tail out to large separations $r_1-r_2\gg 1$. Again we find large 
differences between the unequal-time and equal-time cases, with up to a factor 100 difference 
for large separations in redshift. 
The highly oscillatory nature of the 
functions is a real effect -- not noise in the calculation (estimated error bars 
from the numerical integration are 
smaller than the linewidth in Figure \ref{qplot}). This highly oscillatory nature is a result of multiple Bessel 
functions in the solution moving in and out of phase as the values of $r$ change. For the case of a constant 
power spectrum (that is not a function of $k$-mode) the equations (\ref{qeq}) have an analytic solution given by 
Lommel's integrals \citep{watson1995treatise,abramowitz1964handbook} which are also 
rederived in \cite{1995MNRAS.272..885F}
\bea 
\label{lommel}
\int_0^K {\rm d}k k^2 j_{\ell}(kr)j_{\ell}(k r')&=&
\left[\frac{\pi}{2(rr')^{1/2}}\right]\left(\frac{K}{r^2-r'^2}\right)\left[rJ_{\ell-1}(Kr)J_{\ell}(Kr')-r_2J_{\ell-1}(Kr')J_{\ell}(Kr)\right]\nonumber\\
\int_0^K {\rm d}k k^2 j_{\ell}(kr)j_{\ell}(k r)&=&
\left(\frac{\pi}{2r}\right)\left(\frac{K^2}{2}\right)\left[(J_{\ell}(Kr))^2-J_{\ell-1}(Kr)J_{\ell+1}(Kr)\right], 
\eea 
where $J$ are Bessel functions of the first kind. We note that Lommels integrals 
do not converge to a delta-function as $K\rightarrow \infty$ (equation \ref{edk}), but become undefined. 
In Figure \ref{lplot} we compare the numerical calculations used in this paper with the analytic solution 
in equations (\ref{lommel}) and find that the difference at least a factor of ten smaller than the amplitude of 
$C^{\delta\delta}(\ell,z,z')$. 

\subsection{Projection Effects}
The final part of equation (\ref{Clexact}) that we consider is the effect 
of the projection kernels $F_A(r,r_1)$. 
In Figure \ref{qplot} and equation (\ref{qeq}) we have already investigated the case that 
$F_A(r,r_1)=\delta^D(r-r_1)$. A further case that has been investigated in \cite{2008PhRvD..78l3506L} 
is the case that 
$|\partial \ln F_A(r,r_1)/\partial r_1| \ll 1$, when the `Limber approximation' can be applied to 
equation (\ref{Clexact}); this is applied in the equal-time case 
when the limits on the $k$-mode integral are $0\leq k\le\infty$, and a Laplace 
transform can be used. We note that in the unequal-time case the derivation of \cite{2008PhRvD..78l3506L} is not applicable 
because the $r_1$ and $r_2$ integrations are not separable.  

\begin{figure}
\begin{tabular}{cc}
\includegraphics[width=0.5\textwidth]{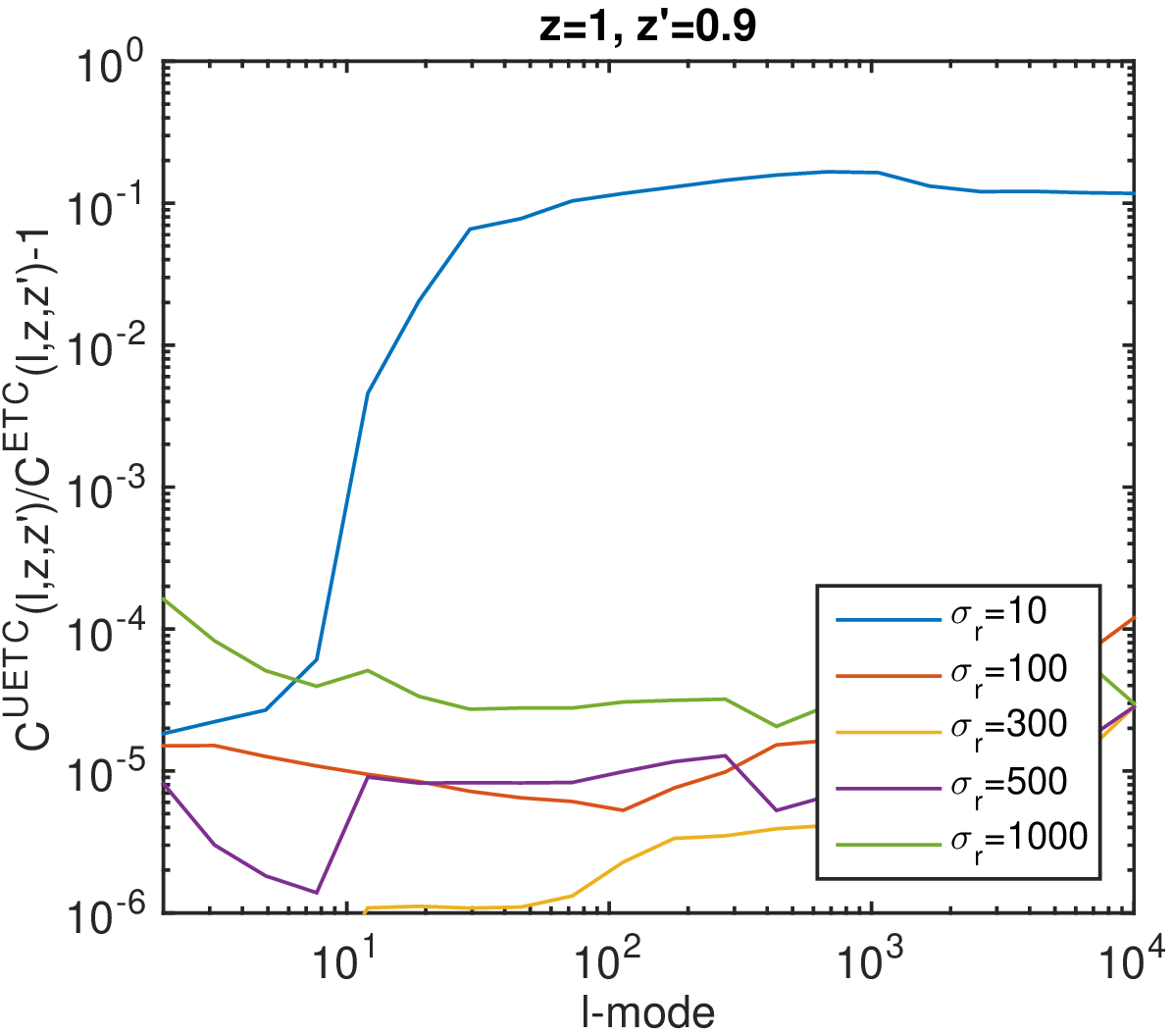}& \includegraphics[width=0.5\textwidth]{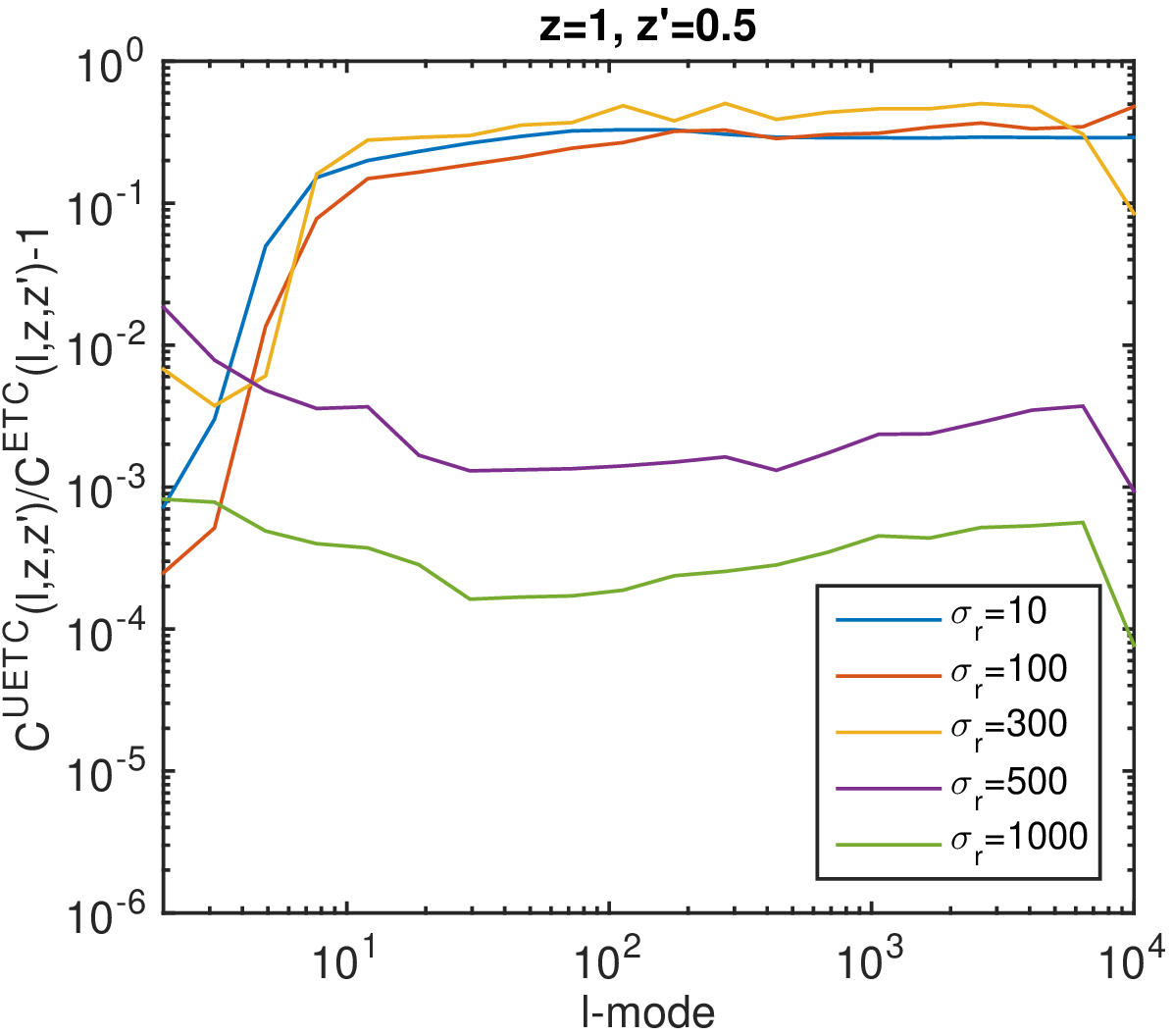}
\end{tabular}
\caption{Left Panel: The fractional change in a projected power spectrum for a Gaussian projection 
kernel $F_A(r,r_1)\propto \exp[-(r-r_1)^2/2\sigma^2_r]$, fixing $F_B(r,r_2)=\delta^D(r-r_2)$. 
We set $r_B=r(z=1)$ and $r_A=r(z'=0.9)$. The lines correspond to different widths of the projection kernel.
Right Panel: Is the same as the left panel except with $r_A=r(z'=0.5)$.}
\label{clplot}
\end{figure}

To investigate the effect of the projection kernels we consider $B$ to be at a fixed redshift:  $F_B(r,r_2)=\delta^D(r-r_2)$ 
and use a Gaussian projection kernel for $F_A(r,r_1)\propto \exp[-(r-r_1)^2/2\sigma^2_r]$. We choose $r_B$ to be the 
comoving distance at $z=1$. We then vary the width of the Gaussian projection kernel $\sigma_r$. In 
Figure \ref{clplot} we show the fractional change in the projected power spectra for two different $r_A$ corresponding
to redshifts of $z=0.5$ and $0.9$. Again the non-smooth nature of these plots is a real feature of the oscillatory nature of the 
unequal-time computations. 

As expected projected cross-correlation power spectra with large redshift separations are affected more by the 
equal-time approximation, which can also be seen in Figures \ref{dplot} and \ref{qplot}. Narrow, well-separated kernels are affected most. Broader kernels integrate over a range of redshifts, including separations where the effect is smaller, and reducing the overall impact. 
We also find that high $\ell$-modes are modified more than small $\ell$-modes, which is consistent with the underlying 
power spectrum being affected most at high $k$ (Figure \ref{pkplot}). For widely-separated bins in redshift the change in 
the projected power spectrum can be over $10\%$ over scales $10^2 \ls \ell \ls 10^4$.
Therefore the cases in cosmology where the equal-time approximation is expected to have greatest impact are those where 
widely-separated tomographic redshift bins are used, with narrow 
projection kernels. 

\section{Application}
\label{Application}
As shown in the previous section the difference between equal-time-approximated projected 
power spectra and the full unequal-time case is expected to become important for cross-correlations 
between widely separated redshift bins of fields that have a sharply-peaked projection kernels. 
An example of such a fields in cosmology are the cross-correlations between tomographic redshift bins 
of the weak lensing intrinsic alignment effect. 

Weak lensing is an integrated effect, whereby the image of a distant galaxy is 
distorted by the gravitational lensing effect caused by matter perturbations along 
a line of sight; and is therefore by nature a projected field. 
In weak lensing the observed ellipticity is (to first order) a sum of the intrinsic (unlensed) ellipticity; 
the additional ellipticity caused by the gravitational lensing effect, $\gamma$, known as shear; and any noise effect: 
$e^{\rm obs}=e^{\rm int}+\gamma+e^{\rm noise}$. When taking the covariance of these quantities and constructing 
projected fields (as in equation \ref{Clexact}) the result is a sum of four terms:
\beq 
C^{\rm Total}_\ell(r, r')=C^{\rm GG}_\ell(r, r')+C^{\rm GI}_\ell(r, r')
+C^{\rm IG}_\ell(r, r')+C^{\rm II}_\ell(r, r'), 
\eeq
where $GG$ refers to the shear-shear correlations, $GI$ to the cross-correlation between the intrinsic ellipticity and 
shear, and $II$ to intrinsic-intrinsic correlations. 
The correlation $IG$ is expected to be zero (or at least very small) but we include it here for completeness. 

The weak lensing observable is proportional to derivatives of the Newtonian potential therefore the 
inner integration over $k$-mode in equation (\ref{Clexact}) is over the power spectrum of the Newtonian potential 
$C^{\Phi\Phi}_{\ell}(r_1,r_2)$,  
\beq 
C^{\rm GG}_\ell(r, r')=\int_0^{r} dr_1 \int_0^{r'} dr_2 F_G(r,r_1) F_G(r',r_2)C^{\Phi\Phi}_{\ell}(r_1,r_2),
\eeq 
for the $GG$ case, where the inner power spectrum is related to the matter power spectrum through Poisson's equations, 
which introduces a factor of $k^{-4}$ so that (up to a cosmology-dependent multiplicative constant)
\beq
\label{phiphi}
C^{\Phi\Phi}_{\ell}(r_1,r_2)=\frac{(\ell+2)!}{(\ell-2)!}\int \frac{2{\rm d}k}{\pi k^2} P(k; r_1, r_2) j_{\ell}(kr_1)j_{\ell}(kr_2); 
\eeq
and similarly for the equal-time case, which can be compared to equations (\ref{qeq}) and (\ref{qeq2}). 
We include a factor of $(\ell+2)!/(\ell-2)!$ here, being the result of  
angular derivatives relating the weak lensing observable (shear) to the projected Newtonian 
potential (see \cite{2005PhRvD..72b3516C, 2016arXiv161104954K}). 

The unlensed observed ellipticity of a galaxy can be affected by the local 
gravitational potential, therefore the 
projection kernel in equation (\ref{Clexact}) is proportional to a 
delta function in this case. The shear effect is 
integrated along the line of sight and the projection kernel is a combination of angular diameter distances that 
determine the geometric gravitational lensing effect. We follow the notation of \cite{2015MNRAS.449.2205K,2016arXiv161104954K} 
and define the power spectra using equation (\ref{Clexact})
where the two fields are combinations of $A=\{G, I\}$ and $B=\{G, I\}$, with projection kernels given by 
\bea 
F_G(r,r_1)F_G(r',r_2)&=&C_G(r) C_G(r') q(r,r_1)q(r',r_2)\nonumber\\
F_I(r,r_1)F_G(r',r_2)&=&C_I(r) C_G(r') p(r,r_1)q(r',r_2)\nonumber\\
F_G(r,r_1)F_I(r',r_2)&=&C_G(r) C_I(r') q(r,r_1)p(r',r_2)\nonumber\\
F_I(r,r_1)F_I(r',r_2)&=&C_I(r) C_I(r') p(r,r_1)p(r',r_2),
\eea 
where $p(r,r_1)$ is the comoving distance probability distribution for galaxies in a bin labelled with $r$. The 
shear kernel is given by 
\beq
q(r,r_1)=\frac{3H_0^2\Omega_{\rm M}}{2c^2}\frac{1}{a(r_1)r_1}\int {\rm d}r' p(r,r')\left(\frac{r'-r_1}{r'}\right)w(r,r'),
\eeq
where $H_0$ is the current Hubble parameter, $\Omega_{\rm M}$ is the current dimensionless matter density, $c$ is the speed 
of light in vacuum, $a(r)$ is the dimensionless scale factor, $w(r,r')=1$ for $r'\leq r$ and zero otherwise; 
here we assume a flat geometry. 
The functions are $C_G(r)=1$ and $C_I(r)=A_{\rm IA}c_1\bar\rho(r)/D(r)(1+z(r))/r^2$; 
where $D(z)$ is the linear growth factor, $\bar\rho(z)$ 
is the mean matter density at redshift $z$, $c_1=5\times 10^{-14}h^{-2}M_{\odot}^{-1}$Mpc$^3$, and $A_{\rm IA}$ is a 
free parameter that has a fiducial value of $A_{\rm IA}=1$. This model for the 
intrinsic alignment power spectrum is the `linear alignment model' discussed in \cite{2004PhRvD..70f3526H} 
(see \cite{2015SSRv..193....1J} for a recent review). 

We use a fiducial Euclid-like\footnote{\url{http://euclid-ec.org}} weak lensing survey \citep{2011arXiv1110.3193L} 
(this is also similar to an LSST survey \cite{2009arXiv0912.0201L}), 
with a survey area of $15,000$ square degrees, 
a surface number density of galaxies of $n_0=30$ per square arcminute, a number density distribution given by 
$n(z)=n_0(3z^2/0.524)\exp[-(z/0.64)^{1.5}]$, and a Gaussian photometric redshift distribution 
with standard deviation $\sigma_z(z)=0.03(1+z)$; the function $p(r,r_1)$ is given by the convolution of the $n(z)$ in each 
tomographic redshift bin and the 
photometric redshift distribution. We use $5$ tomographic redshift bins with central redshifts given by  
$\{0.10, 0.57, 1.04, 1.51, 1.97\}$. We compute these power spectra for the unequal-time and equal-time cases where the 
underlying power spectra are given in equations (\ref{puetc}) and (\ref{petc}) where for the equal-time 
case we use the ansatz $[P(k; r, r')]^2=P^{\rm ETC}(k; r)P^{\rm ETC}(k; r')$. 
\begin{figure}
\includegraphics[width=1.0\textwidth]{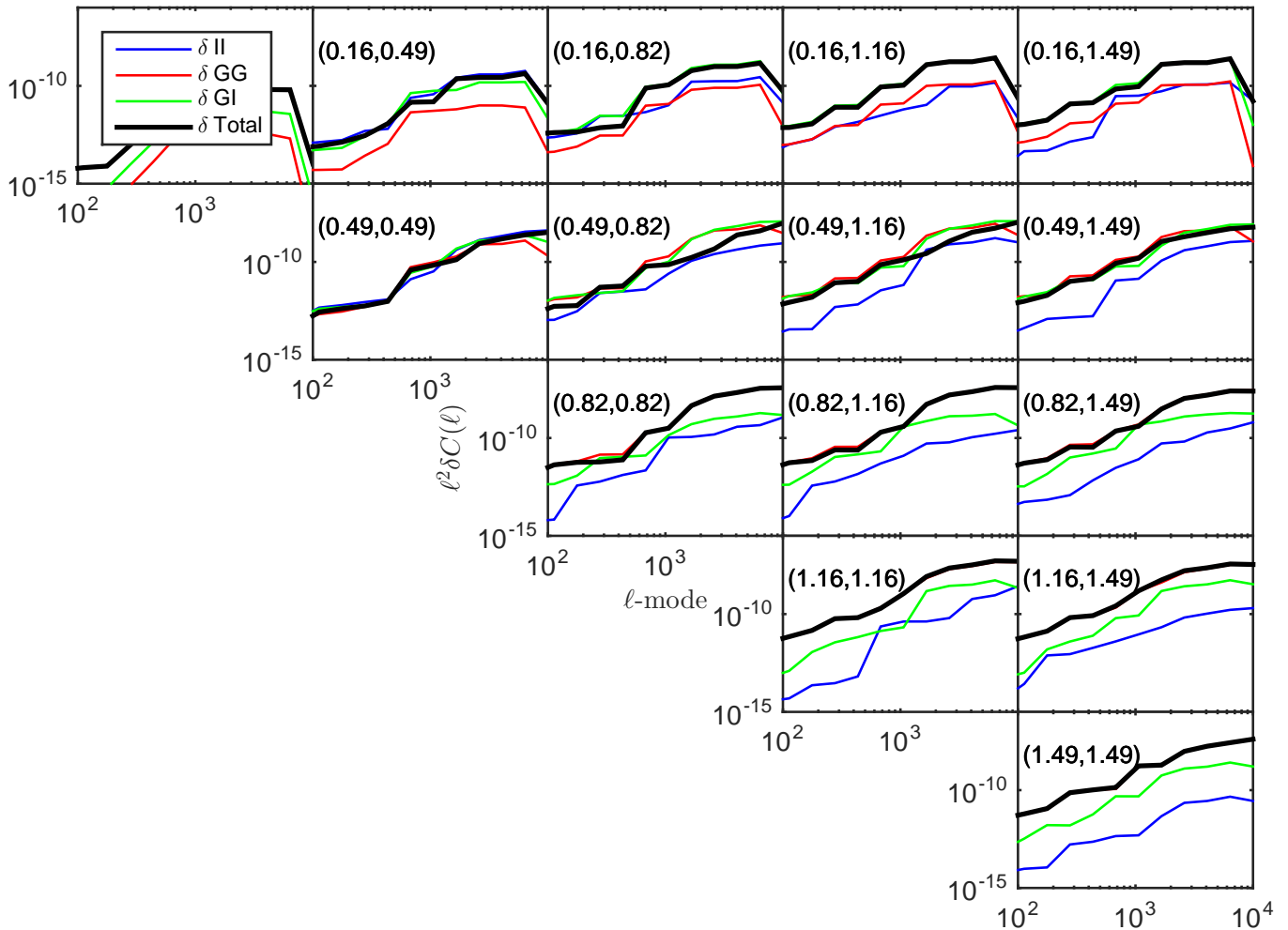}
\caption{The difference $\ell^2 |C^{\rm UETC}_\ell(z,z')-C^{\rm ETC}_\ell(z,z')|$ 
between the unequal-time projected power spectra and the equal-time projected 
power spectra as a function of $\ell$-mode for each redshift bin combination for the $GG$, $GI$, and $II$ cases.}
\label{dclplot}
\end{figure}

In Figure \ref{dclplot} we show the difference between the projected power spectra computed in the unequal-time and 
equal-time cases
\beq 
\delta C_\ell(z,z')=|C^{\rm UETC}_\ell(z,z')-C^{\rm ETC}_\ell(z,z')|. 
\eeq
We find that the $GI$ term is most affected since its amplitude is relatively large and it requires a narrow kernel. The $II$ 
term has a combination of two narrow kernels but its amplitude is smaller relative to the total power spectrum. The $GG$ 
term is also affected but at a typically lower amplitude than the $GI$ term. 
If narrower redshift bins were used then the amplitude of the difference should increase. 
The change in power spectrum is certainly small, but in fact the requirements on $\delta C_\ell(z,z')$ are particularly stringent for 
weak lensing surveys. We can assess the impact of these changes by computing the integrated effect over the differences 
\beq 
\label{barA}
\langle {\mathcal A}\rangle=\frac{1}{2\pi}\frac{1}{N_{\rm bins}}\sum_{z\dot{\rm bins}}\int {\rm d}\ln\ell\,\ell^2 \delta C_\ell(z,z'); 
\eeq
complementary formulations are provide for this quantity in \cite{2013MNRAS.429..661M,2013MNRAS.431.3103C,2008MNRAS.391..228A}, 
the sum is over all redshift bin pairs and $N_{\rm bins}$ is the total number of redshift bin pairs.
In general a non-zero $\langle {\mathcal A}\rangle$ 
will change the amplitude of the power spectrum and bias cosmological parameter inference. 
As discussed in \cite{2013MNRAS.431.3103C} the requirement on the amplitude of this 
quantity is $\langle {\mathcal A}\rangle\leq 10^{-7}$ for a Euclid- or LSST-like weak lensing survey to return 
unbiased results on the dark energy equation of state parameters, 
this requirement is an allowance for \emph{all} systematic effects including instrumental and algorithmic quantities.  
For our fiducial set we find that $\langle {\mathcal A}\rangle =7.4\times 10^{-8}$, which would account for $\simeq 74\%$ 
of the requirement\footnote{Using the more conservative definition for 
equation (\ref{barA}) in \cite{2013MNRAS.429..661M}, 
that includes a normalisation $N_A=(1/2\pi)(1/N_{\rm bins})\sum_{z\dot{\rm bins}}\int {\rm d}\ln\ell\,\ell^2$ 
we find that $\langle {\mathcal A}\rangle/N_A =4.2\times 10^{-13}$, which would account for $23\%$ of the requirement.}. 

\begin{figure}
\includegraphics[width=0.5\textwidth]{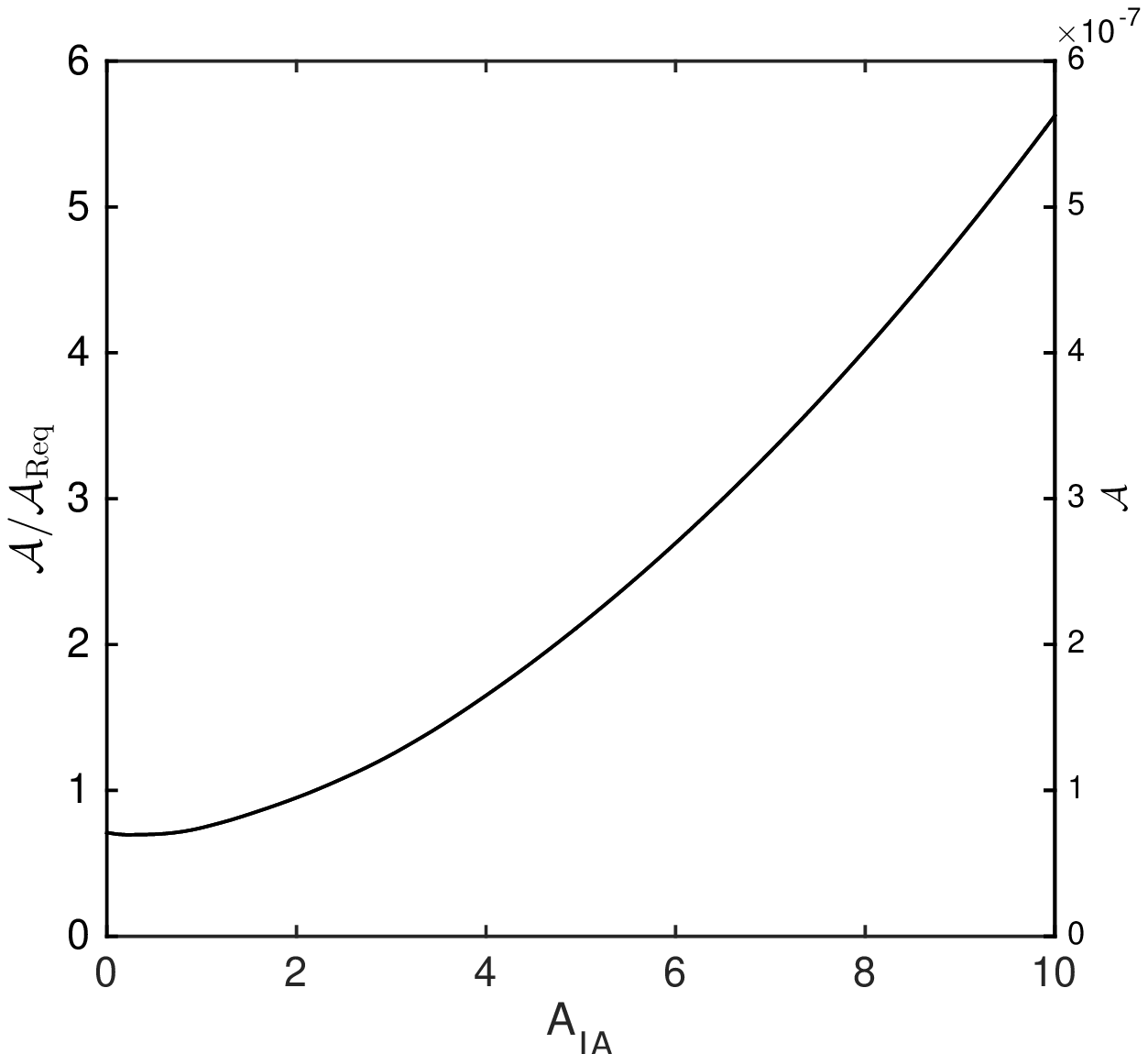}
\caption{The change in the integrated systematic power spectrum, equation (\ref{barA}), caused by the equal time 
assumption, as a function of the intrinsic alignment amplitude $A_{\rm IA}$. 
The left axes we show this normalised by the required value of $10^{-7}$, and the right axes shows 
the unnormalised value.}
\label{aiaplot}
\end{figure}

Because the equal time ansatz is expected to affect projected power spectra with narrow kernels in redshift, 
and the intrinsic alignment power spectrum is such a power spectrum, we investigate the impact of changing the 
amplitude of intrinsic alignment signal. In Figure \ref{aiaplot} we change the parameter $A_{\rm IA}$, that has a fiducial 
value of $A_{\rm IA}=1$ and show how the integrated change in the power spectrum, $\langle {\mathcal A}\rangle$ varies. In 
the left axes we show this normalised by the required value of $10^{-7}$, and the right axes shows the unnormalised value. 
The current range in this parameter is approximately $-1\ls A_{\rm IA}\ls 10$ \cite{2013MNRAS.432.2433H}. We 
find as expected that the the impact of the equal time assumption becomes more prominent as the amplitude of the 
intrinsic alignment signal increases. 

These results serve as an example of the type of impact that the assumption of equal-time correlators may 
have in cosmology, however the current precision of the perturbative approach to representing the matter power spectrum 
over a large range of $k$-modes, and the lack of good predictions for how the unequal-time power spectrum behaves as 
a function of cosmological parameters, mean that the precise numerical results are only indicative of the full 
impact which requires further investigation. 

\section{Conclusion}
\label{Conclusion}
In this paper we have presented a full projected field formalism for cosmology that includes the integration over 
an unequal-time correlator of the matter overdensity power spectrum $P(k; r_1, r_2)$. 
This relaxes an assumption that has been made in cosmology to date where individual or mixed equal-time correlators of the matter 
overdensity have been used, either and ansatz $P(k; r_1, r_2)\simeq [P(k; r_1) P(k; r_2]^{1/2}$ \cite{2005PhRvD..72b3516C}, or 
alternatively simply $P(k; r_1)$ or $P(k; r_2)$.

We investigated the impact of the assumption of equal-time correlators by expanding both $P(k; r_1, r_2)$ and 
$P(k; r_1)$ perturbatively. We find that the impact of this assumption is largest at large separations 
in redshift $|z-z'|\gs 0.3$ where the change in the amplitude of the matter power spectrum can be as much as 
$5-10\%$ for $k\gs 5h$Mpc$^{-1}$. For projected fields $C_\ell(z, z')$ we find that the impact is 
dependent on the width of the projection kernel. For 
Gaussian kernels with full-width-half-maxima of $\sigma_r\ls 300 h^{-1}\,$Mpc we find a $\sim 10\%$ effect for large 
redshift separations $|z-z'|\simeq 0.5$. Therefore observed power spectra that have narrow projections kernels 
and cross-correlations between widely-separated redshift bins are expected to be affected by the equal-time approximation.

One application where there are narrow projected kernels and widely separated redshift bins is the computation of 
weak lensing two-point correlation functions and power spectra -- where in particular the contribution from intrinsic 
alignments (unlensed ellipticities) has local projection kernels in comoving coordinates. We find that 
for a Euclid- or LSST-like weak lensing experiment the assumption of equal-time correlators could introduce 
biases in the measurements of cosmological parameters, and that this is strongly dependent on the amplitude of the 
intrinsic alignment signal. However, due to the lack of good models for how cosmological parameters affect 
unequal-time correlators, these results are indicative only and require further investigation.  

Unequal-time correlators have been investigated previously in 
\cite{1998PhRvL..81.2008A,1997PhRvL..79.1611P,1997PhRvL..79.4736A,1999PhRvD..59b3508A} in the study of cosmic strings. 
To determine unequal-time correlations analytically is an involved calculation, and in particular
for the matter overdensity field perturbation theory at high $k$-mode would make a full analytic solution challenging. 
In this paper we compute the matter power spectrum using perturbation theory to third order and use $k$-modes with 
$k\leq 15h$Mpc$^{-1}$. One approach to include unequal-time correlators in cosmological statistics will be 
to extend this approach for arbitrary cosmological parameters and to high precision; 
such work that has begun in studies for equal-time correlators e.g. \cite{2010PhRvD..81b3524S,2014JCAP...11..039B}.
As noted by \cite{2012PhRvD..85f3509B} the eikonal phase representation of matter overdensity perturbations 
results in equal-time correlations being unaffected 
but unequal-time correlations being damped -- this is a consistent with the results we find here where 
the equal-time case would require a lower matter power spectrum amplitude to fit the same data. 
Another approach will be to determine the 
unequal-time power spectra directly from simulations of the matter field such 
as \cite{2015A&C....13...12N,2015MNRAS.446..521S}. 

When designing statistics in cosmology great care must be taken to investigate the impact of any 
assumption or approximation used. We find that the equal-time approximation can potentially have a large impact on 
cosmological parameter inference. Unequal-time correlators can be computed using either perturbation to high $k$-modes 
or from simulations, both of which require attention in order for future cosmological results to be unbiased.

\begin{acknowledgments}
TDK is supported by a Royal Society University Research Fellowship. We thank E. Komatsu for making his perturbatively 
computed power spectrum code public. 
\end{acknowledgments}

\nocite{*}

\bibliography{apssamp}

\end{document}